\documentclass[%
 reprint,
 amsmath,amssymb,
prb,
]{revtex4-2}

\usepackage{graphicx}
\usepackage{dcolumn}
\usepackage{bm}
\usepackage[version=3]{mhchem} 

\begin{document}


\title{Assessing the role of quantum effects in 2D heterophase \ce{MoTe2} field effect transistors.}

\author{Line Jelver}
\email{lije@mci.sdu.dk}
\affiliation{CAMD, Dept. of Physics, Technical University of Denmark, Bldg. 309, DK-2800 Kongens Lyngby, Denmark.}
\affiliation{Synopsys QuantumATK, Fruebjergvej 3, PostBox 4, DK-2100 Copenhagen, Denmark.}
\affiliation{Center for Nano Optics, University of Southern Denmark, DK-5230 Odense M, Denmark.}

\author{Ole Hansen}
\affiliation{DTU Nanolab, Technical University of Denmark, Bldg. 347, DK-2800 Kongens Lyngby, Denmark.}

\author{Karsten Wedel Jacobsen}
\affiliation{CAMD, Dept. of Physics, Technical University of Denmark, Bldg. 309, DK-2800 Kongens Lyngby, Denmark.}

\date{\today}

\begin{abstract}
The two-dimensional transition metal dichalcogenides (TMDs) have been proposed as candidates for the channel material in future field effect transistor designs. The heterophase design which utilizes the metallic T- or T’ phase of the TMD as contacts to the semiconducting H phase channel has shown promising results in terms of bringing down the contact resistance of the device. In this work, we use ab-initio calculations to demonstrate how atomic-scale and quantum effects influence the ballistic transport properties in such heterophase transistors with channel lengths up to 20 nm. We investigate how the charge transfer depends on the carrier density both in T’-H \ce{MoTe2} Schottky contacts and planar T’-H-T’ \ce{MoTe2} transistors. We find that the size of the Schottky barrier and the charge transfer is dominated by the local atomic arrangements at the interface and the doping level. Furthermore, two types of quantum states have a large influence on the charge transport; interface states and standing waves in the semiconductor due to quantum confinement. We find that the latter can be associated with rises in the current by more than an order of magnitude due to resonant tunneling. Our results demonstrate the quantum mechanical nature of these 2D transistors and highlight several challenges and possible solutions for achieving a competitive performance of such devices.
\end{abstract}

\maketitle


Field effects transistors (FETs) have through several decades been continuously optimized and down-scaled. Today, quantum effects have become one of the limiting factors in the device performance. Tunneling effects between the source and drain as well as through the gate stack make it challenging to continue the down-scaling. Further development of the energy efficiency and computational power is therefore to a larger and larger degree achieved by optimizing the transistor design and materials.\cite{IRDS2020}

The 2D transition metal dichalcogenides (TMDs) represent a promising group of materials which could substitute silicon as the semiconducting channel of the device. These two dimensional semiconductors have the advantage that they, in contrast to silicon, can be synthesized down to the monolayer limit while retaining a high mobility.\cite{Huyghebaert,Akinwande2019} This allows for the ultimate down-scaling of the transistor channel with a typical monolayer thickness of only 0.7 nm. However, two large challenges need to be resolved before these materials can become a viable choice for future transistor designs. Firstly, a scalable method for area selective doping of 2D materials is not currently available.\cite{Balasubramaniam2019} This means that the carrier density in a 2D semiconductor cannot be controlled in a reproducible manner. Secondly, the interface between the 2D channel and the 3D source and drain contacts is difficult to fabricate. Metal deposition on a 2D material can be very destructive and impurities and defects in the interface result in large contact resistances.\cite{Allain2015,Kim2017} A promising solution to this is to exploit that the TMDs exists in different phases which exhibit either metallic or semiconducting properties. The group-IV TMDs are generally most stable in the semiconducting 2H phase but in \ce{MoTe2}, the metallic 1T' phase is meta-stable\cite{Duerloo2014} and can be selectively grown\cite{Yoo2017,Sung2017,Ma2019,Zhang2019,Zhang2019apr,Xu2019,Xu2019sep} or induced from a 2H sample.\cite{Cho2015,Song2016} The \ce{MoTe2} heterophase transistor design relies on this fact and utilizes the 1T' phase of the TMD as the source and drain contact. This approach reduces the contact resistance significantly since charge transport across the interface between 1T'-\ce{MoTe2} and a 3D metal contact seems to be less vulnerable to impurities and defects.\cite{Cho2015,Sung2017,Xu2019,Xu2019sep,Zhang2019,Zhang2019apr,Ma2019} The ultimate prospect of this design is to combine the monolayer versions of these bulk phases, the H- and T' phase, to create a completely two-dimensional transistor.

In our previous work\cite{Jelver2020}, we have used ab-initio calculations to demonstrate how interface resonances at monolayer (ML) T'-H \ce{MoTe2} contacts create peaks in the transmission such that the Schottky barrier heights extracted by the thermionic emission model yield significantly different results than the Schottky barrier seen in the density of states. In this work, we go a step further and investigate the consequences of such quantum effects not only in the Schottky contacts but also in T'-H-T' \ce{MoTe2} transistors.

The investigations are conducted in two steps. The first step is to examine the Schottky barriers and charge transport in two different T'-H \ce{MoTe2} contacts as the carrier density is gradually increased. FETs operate by adjusting an electrostatic field using the gate electrode to draw in or push out carriers in the semiconductor channel. The same effect can therefore be created by gradually increasing or decreasing the carrier density in the semiconductor at a single metal-semiconductor interface. Previous ab-initio studies by Saha \emph{et al.}\cite{Saha2017} and Urquiza \emph{et al.}\cite{Urquiza2020} have investigated two different doping densities of T'-H \ce{MoS2} Schottky contacts and concluded that an increased doping density decreases both the barrier height and the depletion width and our previous work\cite{Jelver2020} has demonstrated the same effect in T'-H \ce{MoTe2} contacts. However, to our knowledge, this is the first time the properties of these Schottky contacts have been investigated by gradually changing the carrier density. This gradual change allows for mimicking the effects of an applied gate potential in a field effect transistor and the calculations therefore provide new fundamental understanding of the electrostatic control over the charge transfer in these 2D contacts.

The second step is to directly study the charge transport in \ce{MoTe2} heterophase FETs. The two contacts investigated in the previous part are combined to form complete 2D FETs with channel lengths of 10, 18, and 20 nm. In case of the 20 nm channel, the self-consistent transport calculations includes 573 atoms which takes this ab-initio approach to the limit of what is computationally practical. Devices on this length scale have not previously been investigated and help to shed light on the behavior of heterophase FETs in a regime where the channel lengths are comparable to channel lengths used in commercial devices.\cite{IRDS2020}

Our analysis of the Schottky contacts demonstrates that the local atomic configuration at the interface completely dominates the Schottky barrier height, depletion width, and charge transfer. Furthermore, we find that quantum effects play an important part in the charge transfer at the heterophase devices. Both in case of the contacts and the transistors, quantum states bridge the Schottky barrier and contribute significantly to the total current. In the contact, interface states increase the transmission probability and lowers the effective Schottky barrier height. In the transistors, standing waves in the channel increase this effect and results in large current fluctuations in the transfer characteristics, even at a channel length of 20 nm's. However, we find that our devices are not competitive with silicon technology neither in terms of efficiency nor output current. The largest challenges for this heterophase design is that the Schottky barrier at the source, which partly determines the efficiency, is controlled by the local atomic configuration and carrier density in the channel and that the ballistic ON-current is inhibited due to the restriction of momentum conservation along the interface.

The paper is organized as follows. In Section \ref{sec:method}, we explain the details regarding the applied methods and the setup of the Schottky contacts and heterophase transistors. In Section \ref{sec:contacts}, we demonstrate how the Schottky contacts respond to increasing the carrier density. This part serves as background to understand the results of the complete transistor simulations which are the subject of Section \ref{sec:fet}. After demonstrating the results of the device simulations, we discuss how these calculations compare to previous experiments and calculations on the \ce{MoTe2} heterophase transistor and how the performance of these devices could be optimized. This is the subject of Section \ref{sec:disc}. Finally, we summarize the main conclusions in Section \ref{sec:conc}.

\section{Method\label{sec:method}}

\begin{figure*}
    \centering
    \includegraphics[width=.69\textwidth]{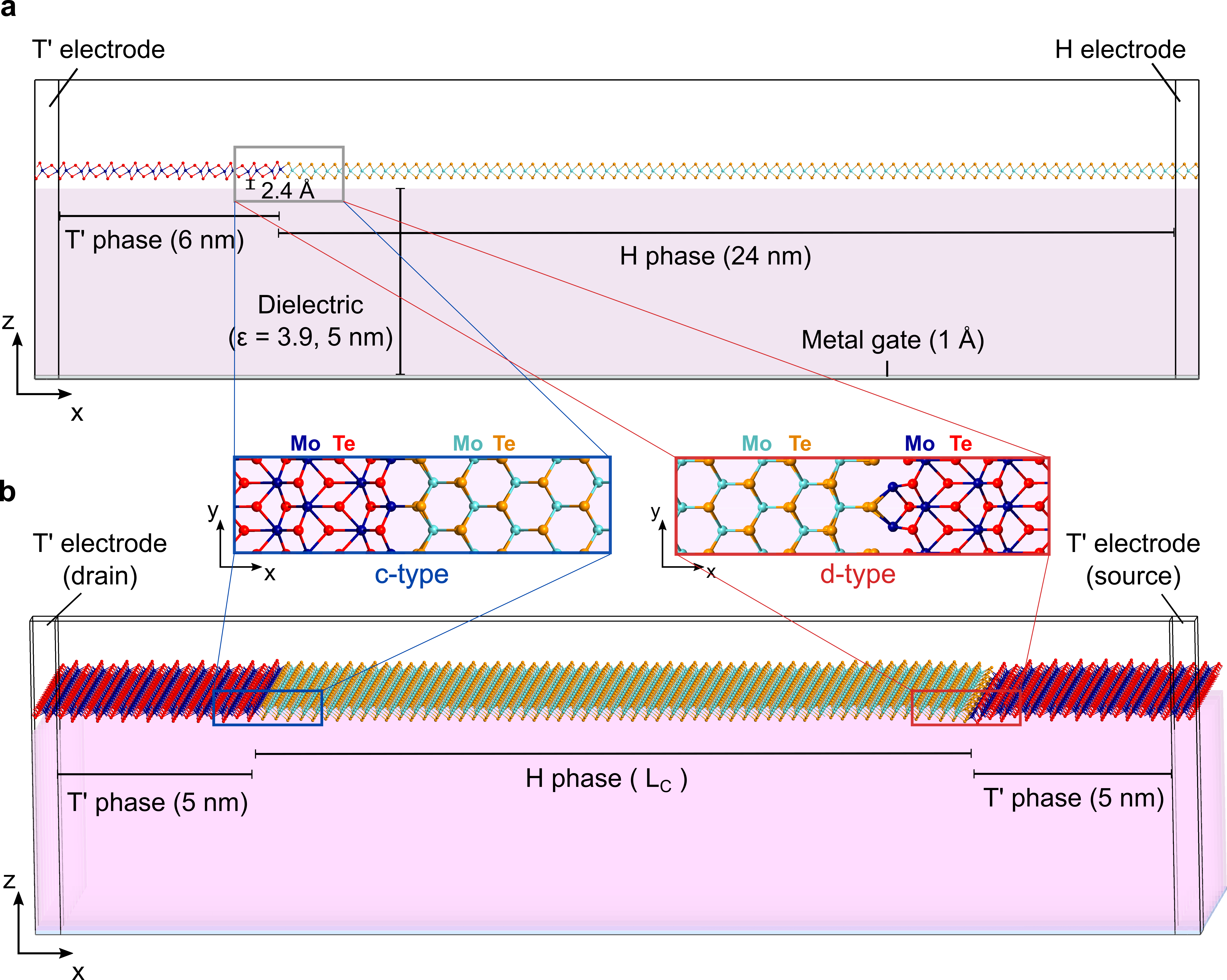}
    \caption{The T'-H Schottky contact and T'-H-T' transistor of monolayer \ce{MoTe2} on a substrate and back-gate. \textbf{a} shows the computational cell of the contacts consisting of a central region and a T'- and H-phase electrode. The total cell size is (30.0, 0.718, 15.0) nm. \textbf{b} shows the computational cell of the 2D transistors with T' phase source and drain and a H phase channel. The total cell size is (10.0+$L_C$, 0.718, 15.0) nm with $L_C$ being 10.0, 18.0 or 20.0 nm. The middle part shows the atomic configurations of the two investigated interface geometries seen from the top.}
    \label{fig:substrate}
\end{figure*}

Interfaces between the two phases of \ce{MoTe2} are most commonly observed along the zigzag direction or at a 60 degree angle to this direction.\cite{Sung2017,Xu2020,Han2020} Here, we choose to investigate interfaces along the zigzag direction. In this direction, the unit cells of the T' and H phase TMD can be combined in 8 different ways while conserving the stoichiometry of the two phases. A stability analysis of these eight interfaces is included in the Supplementary Information.$^\dag$ The c- and d-type are chosen, since these represent some of the most stable interfaces, the c-type in Te-rich conditions and the d-type in intermediate and Mo-rich conditions, meanwhile demonstrating the large variance in interface dipole represented by the different interface geometries.

The c- and d-type interfaces are shown in the middle part of Figure \ref{fig:substrate}. Both of these interfaces have been found by combining the two unit cells along the zigzag direction and thereby creating a strain of 2.12 \% in the y-direction of the T' phase. The three unit cells of each phase closest to the interface have been relaxed to a force tolerance of 0.02 eV/\AA. The unit cell is doubled in the y-direction compared to the conventional unit cells to allow for a stabilizing distortion at the interface. This distortion is clearly visible at the d-type interface.

We use density functional theory (DFT)\cite{DFT1,DFT2} with the Perdew-Burke-Ernzerhof (PBE) exchange-correlation functional\cite{PBE} and the non-equilibrium Green's function (NEGF) method as implemented in QuantumATK.\cite{Stokbro2019} The wave functions are expanded as linear combinations of atomic orbitals using PseudoDojo pseudopotentials\cite{VanSetten2018} using a cut-off energy of 100 Ha. The method is described in details in our previous work\cite{Jelver2020} on similar interfaces where we also justify using PBE to describe these systems. Within this description, the H phase is a direct bandgap semiconductor with a band gap of 1.03 eV and the T' phase is gapless. The computational cell of the Schottky contacts is shown on Figure \ref{fig:substrate}a. The total cell size of the central region is (30.0, 0.718, 15.0) nm and the k-space grid of the NEGF calculations is (401, 6, 1). Figure \ref{fig:substrate}b shows the computational cell of the 2D planar FETs which combines a c- and d-type interface as the drain and source contacts, respectively. The symmetry of the H phase inhibits using the same interface geometry at both source and drain which is why the device is a combination of the two different geometries. The cell sizes of the central region is (10.0+$L_C$, 0.718, 15.0) nm where $L_C$ is either 10.0, 18.0 or 20.0 nm and the k-space grids are equivalent to the Schottky contact calculations.

To simulate the electrostatic environment in a FET, both the 2D Schottky contacts and transistors have been placed on top of a thin metal back gate and a dielectric region of 5 nm with a dielectric constant of 3.9 corresponding to \ce{SiO2}. Both are described as a classical continuum. The lowest atom of the system is placed 2.4 \AA \ above the dielectric region. This value is chosen since it lies close to the experimental value of the distance between 2H \ce{MoS2} and \ce{SiO2}\cite{PhysRevB.13.3843} and since we have found that a variation between 2.4 and 4 \AA \ does not change the barrier height or depletion width. The electrostatic effect of the substrate is to screen the interface dipole which increases the charge transfer since the potential difference between the metal and semiconductor is conserved despite the presence of the substrate. This increases the depletion width but conserves the barrier height. 

We use a continuous doping model to add charge carriers to the semiconductor and we apply a p-type doping since \ce{MoTe2} most commonly exhibit a p-type intrinsic doping in ambient conditions.\cite{Sung2017,Wang2018} The electrons per atom are explicitly modified by scaling the atomic densities.\cite{doping} We choose to dope all the atoms in the system to avoid any artificial effects of defining a sharp interface between the two phases. The extra holes in the metallic phase have a negligible effect.

\section{Schottky contacts\label{sec:contacts}}

\begin{figure}
    \centering
    \includegraphics[width=.99\columnwidth]{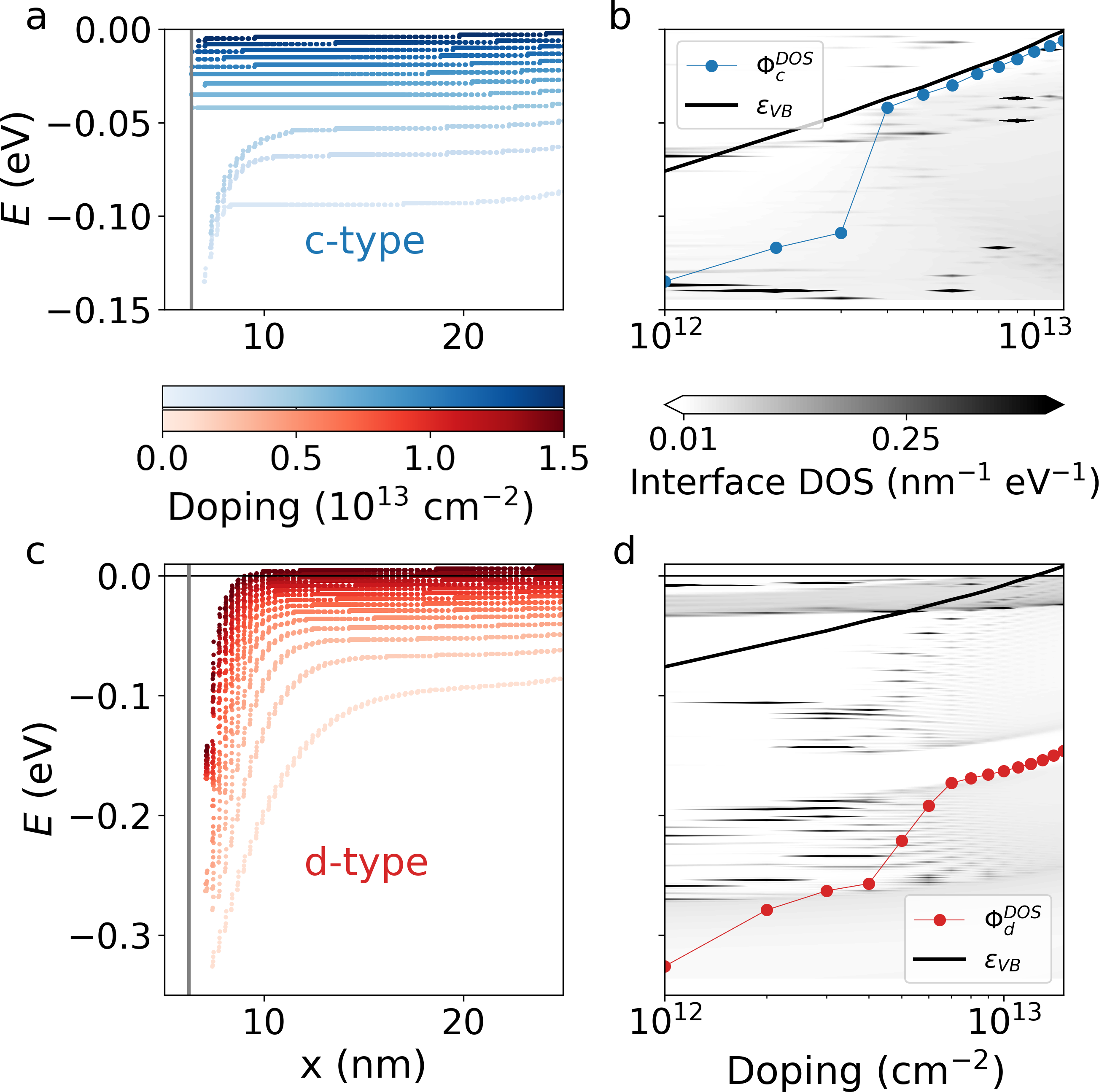}
    \caption{The Schottky barrier evolution with varying levels of p-doping. \textbf{a} and \textbf{c} show the valence band edge of the H phase for the c- and d-type contacts, respectively. The grey line indicates the first Mo-atom belonging to the H phase. \textbf{b} and \textbf{d} show the barrier heights as a function of the doping level (circles) along with the maximum of the valence band (black line). The density of states at the position of the grey line on \textbf{a} and \textbf{c} at each doping level are included as a grey-scale contour to illustrate the density of interface states between the Fermi level and valence band maximum.}
    \label{fig:doping}
\end{figure}

Our starting point is to investigate the two Schottky contacts. The results from these investigations will help to analyze the results of the simulations on the complete transistor setup. The p-doping level is varied between $10^{12}$ and $1.5 \times 10^{13}$ cm$^{-2}$ and the effects on the Schottky barrier height, depletion width, and charge transport are investigated. The Schottky barrier height is found from the projected density of states (DOS) as the distance from the Fermi level to the minima of the valence band. The valence band edge is found by defining the gap region as energies at which the projected DOS is negligible. The evolution of the valence band positions of the two contacts, when the doping is increased, are shown on Figure \ref{fig:doping}a and \ref{fig:doping}c. The Fermi level is placed at the energy zero point and the x-coordinate of the leftmost Mo-atom, which belonged to the H-phase before relaxation, is marked by a vertical grey line. The c-type contact only shows band bending for the three lowest doping levels. From then on, the valence bands are flat and the barrier height coincides with the maximum of the valence band. The d-type contact show band bending for all the investigated doping levels.

\begin{figure}
    \centering
    \includegraphics[width=.85\columnwidth]{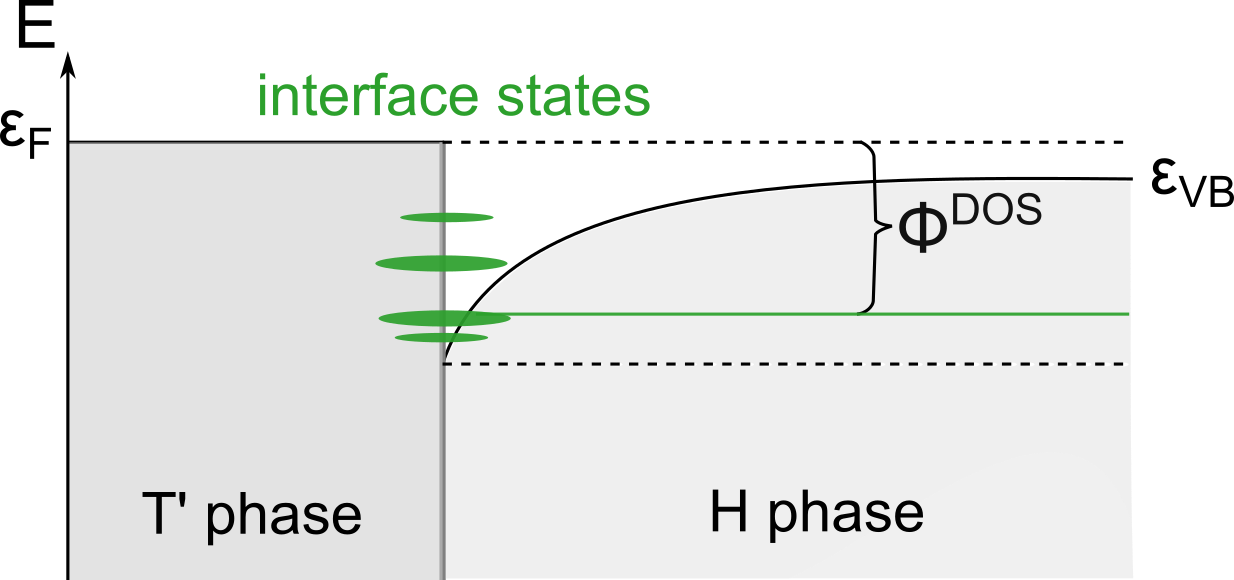}
    \caption{The effect of interface states on the DOS barrier, $\Phi^{DOS}$, between H- and T' phase \ce{MoTe2}. The electrostatic field from the interface dipole causes the valence band to bend downwards creating a barrier height represented by the dashed black line. The DOS associated with interface states bridges this electrostatically defined barrier reducing the effective barrier for quantum transport illustrated by the green line.}
    \label{fig:sketch}
\end{figure}

The barrier height is shown together with the valence band maxima as a function of the doping level on Figure \ref{fig:doping}b and \ref{fig:doping}d. The size of the barrier differs significantly between the two interface geometries and so does the behavior with the doping density. The barrier height does not vary smoothly with the doping level but rather jumps between energy plateaus. Our previous work\cite{Jelver2020} and the work by Liu \emph{et al.}\cite{Liu2018} have demonstrated that these heterophase \ce{MoTe2} contacts host interface states which increases the tunneling probability. At the interfaces investigated here, some of these states bridge the Schottky barrier and result in both increased tunneling probability and an effectively lower barrier than what the strength of the interface dipole dictates. This effect is illustrated on Figure \ref{fig:sketch}. The interface dipole is created due to the depletion of holes in the valence band and the corresponding charge accumulation in the metal which screens the depletion charge. The band bending therefore reflects the strength of the interface dipole and is an electrostatic effect. The density of states, on the other hand, is a representation of the available quantum states which electrons and holes can occupy. The barrier height extracted from the density of states therefore represents an effective barrier for quantum transport which does not necessarily coincide with the electrostatic barrier. This effective barrier will in the following be referred to as the DOS barrier, $\Phi^{DOS}$, in agreement with the definition in our previous work.\cite{Jelver2020}

The density of interface states have been included on Figure \ref{fig:doping}b and \ref{fig:doping}d as a contour plot. The contour represent the DOS at the position of the vertical grey line on Figure \ref{fig:doping}a and \ref{fig:doping}c. The states of interest lie at energies between the bottom of the barrier and the maximum of the valence band. Energies below this range host the valence band states and energies above cannot contribute to transport since these lie within the band gap of the H phase electrode. The effect of the interface DOS is most easily seen in the case of the d-type contact. The barrier bottom is seen to jump between windows of low interface DOS around -0.17 and -0.26 eV. Note, that the slope of the interface states on this plot are much lower than the slope of the valence band maxima. This demonstrates that the interface states originate from the T'-phase rather than the H-phase which was the case for the interface resonances in our previous study as well. The interface DOS illustrated on the contour plot can be associated both with regular metal induced gap states, i.e. the exponential tail of bulk states in the metal, and such interface resonances.

\begin{figure}
    \centering
    \includegraphics[width=.99\columnwidth]{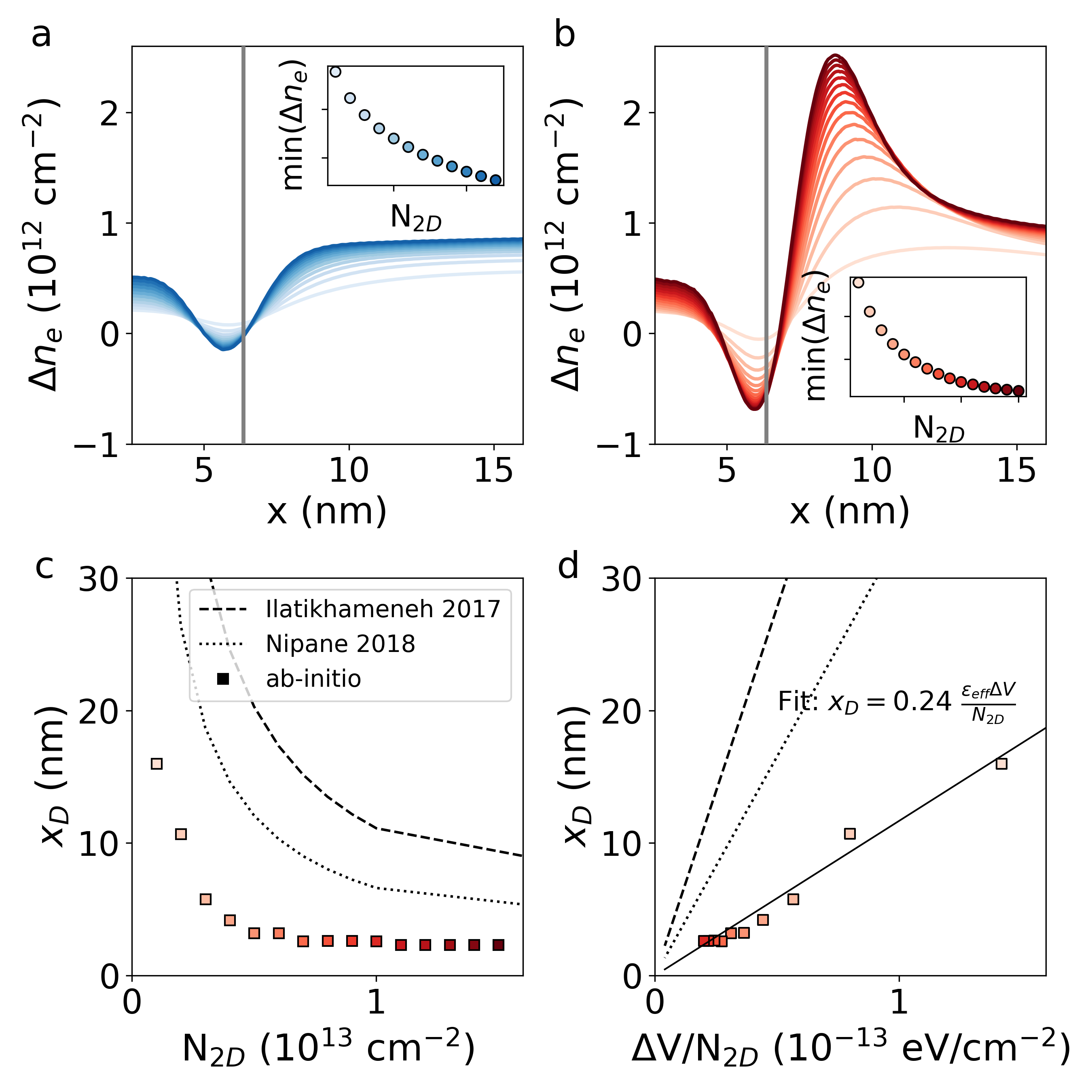}
    \caption{The electron density at the two Schottky contacts (\textbf{a} and \textbf{b}) and the depletion width (\textbf{c} and \textbf{d}) of the d-type interface with varying p-doping following the color-scale on Figure \ref{fig:doping}. The depletion width is compared to the predictions by Ilatikhameneh \emph{et al.}\cite{Ilatikhameneh2017} (dashed line) and Nipane \emph{et al.}\cite{Nipane2017} (dotted line). \textbf{d} shows the depletion width as a function of the work-function difference divided by the doping density and a linear fit to the calculated depletion widths (solid line).}
    \label{fig:depletion}
\end{figure}

These results illustrate that the barrier height depends non-trivially on the doping level due to the presence of interface states. We will now investigate how the interface dipole and depletion width vary with the doping level. In these monolayer contacts, the electronic density and the potential depend a lot on the atomic rearrangements at the phase boundary. To visualize how the interface dipole changes with the doping level, the electronic density of a contact with a low doping level, $N_{2D}^{ref} = 1\times10^{11}$ cm$^{-2}$, is used as a reference. Almost no charge transfer is possible in the reference system since less than 0.02 holes are available in the entire semiconductor region. After the subtraction of the reference density, the doping difference between the reference device and this device is subtracted in order to remove the background doping density,

\begin{align}\nonumber
    \Delta n_e(x) &= \frac{1}{W} \int [ n_e(x,y,z) - n_e^{ref}(x,y,z) ] \, \mathrm{d}y \, \mathrm{d}z \\
    &- (N_{2D}^{dop} - N_{2D}^{ref}).
\end{align}

This difference in electronic density can be seen on Figure \ref{fig:doping}a and \ref{fig:doping}b for the c-type and d-type contacts, respectively. The difference in the size of the dipole is quite clear. The c-type contact shows much less variation of the electronic density through the device which is a footprint of a very small charge accumulation in the depletion region and corespondingly a small barrier height. The d-type contact shows a much larger variation and the interface dipole is seen to become more narrow in space as the doping is increased. The minimum of the density can be used to illustrate the smooth variation of the electronic density with the doping level and is shown on the insets.

Since the depletion width vanish already at a doping level of $4\times10^{12}$ cm$^{-2}$ in the c-type contact, we only consider how the depletion width varies with the doping level of the d-type contact. The depletion width can be found from the bending of the valence bands on Figure \ref{fig:doping}c. The values are compared on Figure \ref{fig:depletion}c and \ref{fig:depletion}d to the expressions derived using classical electrostatics by Ilatikhameneh \emph{et al.}\cite{Ilatikhameneh2017} and Nipane \emph{et al.}\cite{Nipane2017},

\begin{align*}
    x_D^{\mathrm{Ilatikhameneh}} &= \frac{\pi \varepsilon_{\mathrm{eff}} \Delta V }{\ln(4)qN_{2D}}, \\
    x_D^{\mathrm{Nipane}} &= \frac{\pi^2 \varepsilon_{\mathrm{eff}} \Delta V }{8GqN_{2D}}.
\end{align*}

$\Delta V$ is the work function difference between the two materials in the junction, $\varepsilon_{\mathrm{eff}}$ is the effective dielectric constant which is determined by the dielectrics surrounding the 2D material which in this case is $\varepsilon_{\mathrm{eff}} = (\varepsilon_{above}+\varepsilon_{below})/2 = 2.45\varepsilon_0$, $G\approx0.915$ is Catalan’s constant, and $q$ is the elementary charge. These depletion widths have been calculated using the work function of the H- and T' phase electrode at each doping level with computational parameters matching those used for the electrodes of the devices.

Comparing to the ab-initio results, it is clear that the depletion width can not be predicted from the difference in workfunctions and doping level alone. First of all, the values of the two different phase boundaries, as already mentioned, differ a lot and secondly, the depletion width of the d-type contact is considerably shorter than predicted by the classical models. However, as demonstrated by Figure \ref{fig:depletion}d, the classically predicted linear scaling with $\Delta V/N_{2D}$ seems to be fulfilled.

\begin{figure}
    \centering
    \includegraphics[width=.99\columnwidth]{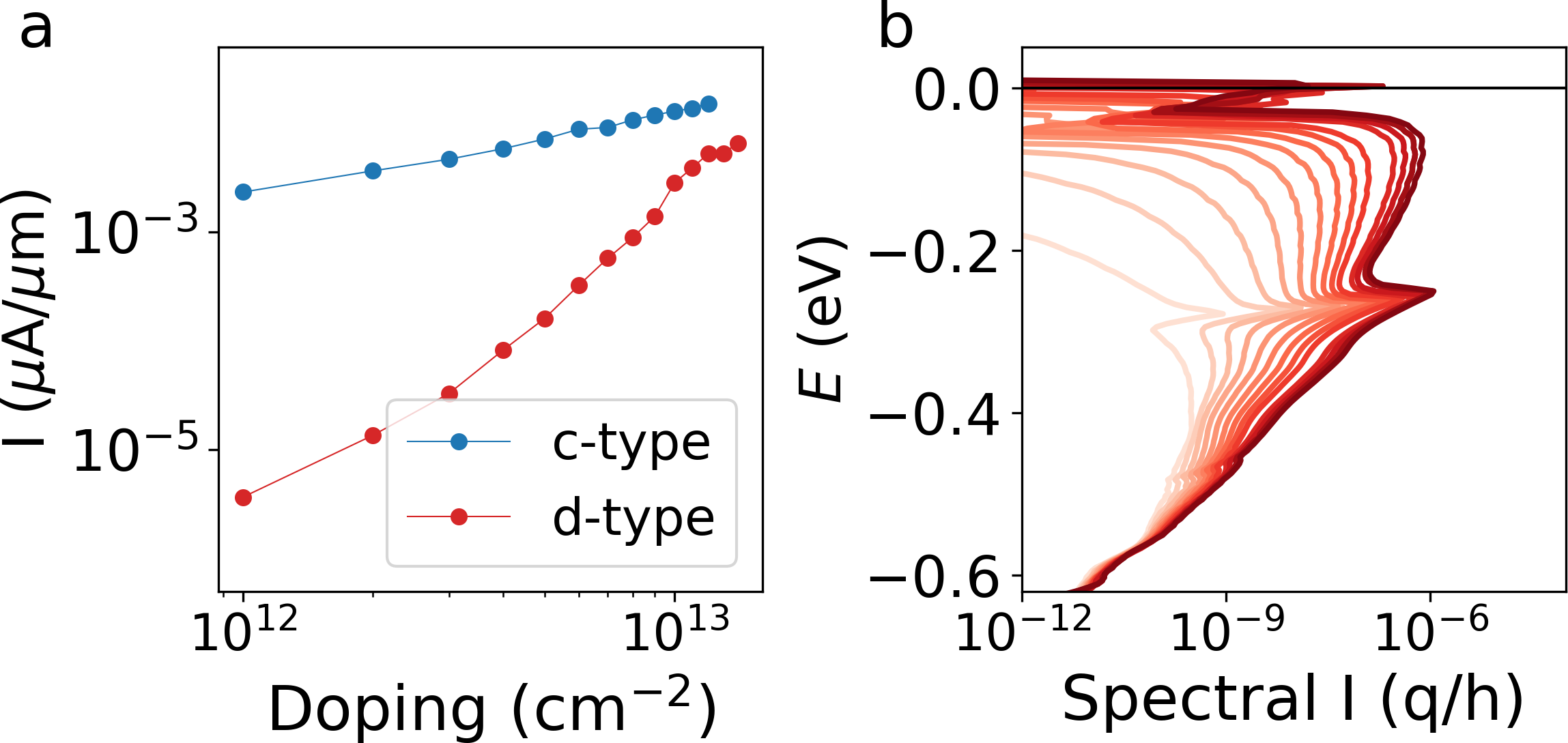}
    \caption{Total and spectral current though the Schottky contacts at varying levels of p-doping. \textbf{a} shows the total current through the two contacts as a function of the doping level and at a bias of $V_{bias}=10$ mV. \textbf{b} shows the spectral current of the d-type contact at varying doping levels. The color-scale follows Figure \ref{fig:doping}.}
    \label{fig:IV}
\end{figure}

Finally, we investigate how the charge transport though the contact changes with the carrier density. We apply a small positive bias of $V_{bias}=10$ mV across the contacts. This drives a hole current from the metal to the semiconductor, the current can be calculated using the Landauer-Büttiker formula,

\begin{align} \nonumber
	I &= \frac{2q}{h} \int T(E,\mu_L,\mu_R) \times \\
	&\left[ f \left( \frac{E-\mu_L}{k_BT} \right) - f \left( \frac{E-\mu_R}{k_BT} \right) \right] \mathrm{d}E. \label{eq:I}
\end{align}
$h$ is Plack's constant, $k_B$ is the Boltzmann constant, $T(E,\mu_L,\mu_R)$ is the transmission, $f$ is the Fermi-Dirac distribution,  $\mu_L$ and $\mu_R$ are the chemical potentials of the two electrodes, defined as $\mu_L-\mu_R=q V_{bias}$, and $T$ is the temperature. We use $T=300$ K to find the current at room temperature. The resulting currents through the two contacts are shown on Figure \ref{fig:IV}a. The current has a relatively smooth dependence on the carrier density and the magnitude differs about two orders of magnitude between the two geometries at low doping levels but reaches a comparable value at high doping levels. Note, that the current is shown on a logarithmic scale. The variation of the spectral current through the d-type contact is shown on Figure \ref{fig:IV}b. The spectral current is the term inside the integral in eq \eqref{eq:I}. Two peaks are present at the energies where a high and relatively constant interface DOS can be identified on Figure \ref{fig:doping}d at all the investigated carrier densities. Those states therefore contribute significantly to the current regardless of the value of the Schottky barrier height and we believe that they can be associated with interface resonances.

To summarize, the Schottky barrier in heterophase \ce{MoTe2} Schottky contacts can be electrostatically tuned to increase the current such that these can be utilized in a transistor. As expected, the interface dipole and charge transfer respond in a smooth manner when increasing the carrier density. This smooth increase is also reflected in the depletion width which scales reciprocally with the carrier density. However, the size of the barrier height, the length of the depletion width, and the total magnitude of the current are determined by atomic-scale and quantum effects. It has been established several times\cite{Li2017,Saha2017,Urquiza2020}, which we also confirm with these calculations, that the barrier height in the TMD heterojunctions depends on the local atomic positions at the phase boundary. From our investigation of the depletion width, we therefore naturally find that the magnitude of this quantity, as well as the size of the interface dipole, is highly dependent on the interface geometry as well. The barrier height behaves non-trivially when the carrier density is increased since interface states can bridge the barrier. Some of these interface states increase the transmission probability significantly and therefore also increase the current trough the contact.

\section{Heterophase FETs\label{sec:fet}}

After concluding these initial investigations of the isolated Schottky contacts, we are ready to simulate a complete transistor setup with source, drain, channel and back-gate as shown on Figure \ref{fig:substrate}b. We combine the two Schottky contacts to create transistors with channel lengths of 10, 18 and 20 nm's and choose a doping level of $N_A=5\times10^{12}$ cm$^{-2}$.

\begin{figure*}
    \centering
    \includegraphics[width=.7\textwidth]{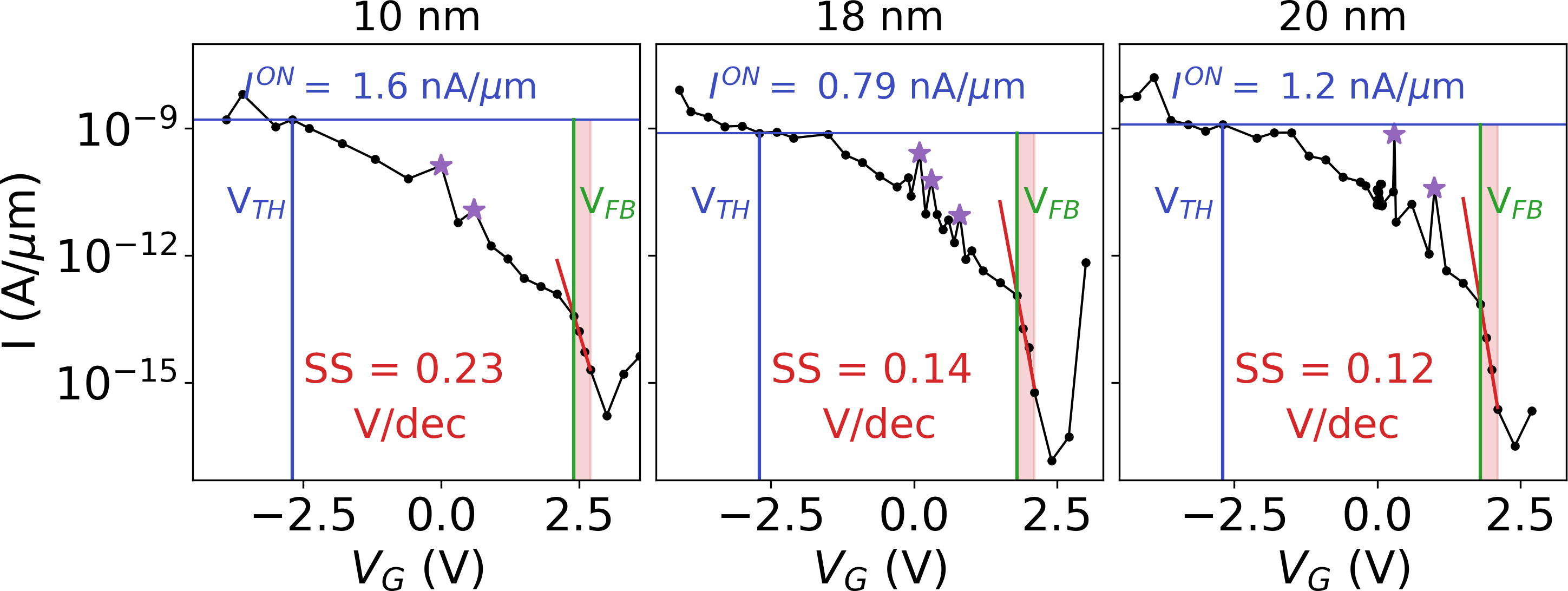}
    \caption{Transfer characteristics of the heterophase FETs at $V_{bias}=-10$ mV. The ON-currents and sub-threshold slopes (SS) are indicated. The SS values are found from the fit indicated by the red lines and the sub-threshold regimes are indicated by the shaded red regions. The flat-band condition, $V_{FB}$, is marked by the green lines and the threshold potential at $V_{TH}=-2.7$ V by the blue lines. Purple stars indicate rises in the current due to resonant tunneling.}
    \label{fig:IVG}
\end{figure*}

\begin{table*}
    \centering
    \small
    \renewcommand{\arraystretch}{1.5} 
    \setlength{\tabcolsep}{4pt}
    \begin{tabular}{cccccccccccc} \hline
    Ref. & $t_{EOT}$  & $t_{ch}$ & $L_C$ & $I^{ON}$ & $V_{sd} (I^{ON})$ & $V_{DD}$ & SS \\ \hline
    \citet{Zhang2019apr} & 1.87 nm & 8 nm & 4 $\mu$m & 0.25 $\mu$A/$\mu$m & 0.1 V & 0.8 V & 69 mV/dec  \\ 
    \citet{Sung2017} & 300 nm  & 5-6 nm  & 4-15 $\mu$m & 15 nA/$\mu$m & 50 mV & 40 V \\
    \citet{Ma2019} & 300 nm & 8 nm  & 20 $\mu$m & 39.6 nA/$\mu$m & -0.1 V & 100 V   \\ \hline
    This Work & 5 nm & ML & 10 nm & 1.6 nA/$\mu$m & -10 mV & 5.7 V & 0.23 V/dec \\
    This Work & 5 nm & ML & 18 nm & 0.79 nA/$\mu$m &  -10 mV & 5.1 V & 0.14 V/dec \\
    This Work & 5 nm & ML & 20 nm & 1.2 nA/$\mu$m &  -10 mV & 5.1 V & 0.12 V/dec \\
    This Work* & 5 nm & ML & 20 nm & 21 nA/$\mu$m  &  -0.1 V & - & - \\ 
    This Work* & 5 nm & ML & 20 nm & 0.15 mA/$\mu$m  &  -0.7 V & - & - \\
    \hline
    IDRS 2022\cite{IRDS2020} & 1 nm & 6 nm\footnote{Fin width of the finFET design.} & 16 nm & 912 $\mu$A/$\mu$m & 0.7 V & 0.7 V & 82 mV/dec &  \\ \hline
    \end{tabular}
    \caption{Equivalent oxide thicknesses, $t_{EOT}$, channel thicknesses, $t_{ch}$, channel lengths, $L_C$, ON-currents, $I^{ON}$, power supply voltages, $V_{DD}$, and sub-threshold slopes (SS) of \ce{MoTe2} based heterophase devices compared to the IRDS 2022 goals.\cite{IRDS2020} *Extrapolated non self-consistently.}
    \label{tab:I_comp}
\end{table*}


The transfer characteristics of the 2D FETs are shown on Figure \ref{fig:IVG} at a bias of $V_{bias}=-10$ mV. This bias results in a device with a d-type barrier at the source and flat bands at the drain. The transfer characteristics are calculated by gradually varying the potential between the back-gate and source electrode and calculate the current using eq. \eqref{eq:I}. At a gate potential of 3.0 V, in case of the 10 nm channel, and 2.4 V, in case of the other two devices, the current is switched completely off. Above those potentials, the devices shift polarity and begin to be dominated by electron transport in the conduction band. The sub-threshold regimes where the current rises exponentially are indicated as the shaded red regions on Figure \ref{fig:IVG}. The sub-threshold slopes in this region are 0.24, 0.14 and 0.12 V/dec for the 10, 18 and 20 nm FETs respectively. The values of the 18 and 20 nm devices agree reasonably well with the SS value of a heterophase \ce{MoS2} device fabricated by Nourbakhsh \emph{et al.}\cite{Nourbakhsh2016} of 120 mV/dec whereas it is considerably larger than the SS value measured in a \ce{MoTe2} heterophase device fabricated by Zhang \emph{et al.}\cite{Zhang2019apr} of merely 69 mV/dec. The impressively low SS value in the latter device is most likely due to the optimized setup used in this case. The heterophase transistor was placed on the high-$\kappa$ dielectric \ce{HfO2} with an equivalent oxide thickness (EOT) of 1.87 nm's and the back gate positioned only below the 2H phase. A low EOT generally both decreases the SS value and increases the ON-current in a device.\cite{Houssa2016}

In the Supplementary Information,$^\dag$ we use a classical electrostatic model to estimate how an increment in the applied gate potential will result in a potential increment at the position of the semiconductor in a device setup such as ours. The relation between the two is given by the ideality factor, $\eta$, which determines the SS value, $\mathrm{SS} = \eta \times 60$ mV/dec. The model assumes a constant charge density in the channel and estimates the effect of the geometry of the device which creates a capacitive potential divider formed by the gate capacitance in series with the capacitance of the substrate and vacuum region above the transistor. This model predicts SS values of 0.20, 0.12, and 0.12 V/dec for the 10, 18 and 20 nm channels respectively. Most of the deviation from the thermodynamic limit of $\mathrm{SS}=60$ mV/dec can therefore be explained by the setup of the device. The remaining discrepancy is most likely due to the variation in the charge density along the channel and the tunneling between the source and drain. The source-drain tunneling is most important in the 10 nm device which explains why the SS value of this device is very large.

The flat-band condition, $V_{FB}$, is met when the exponential dependence of the current breaks off. This point is marked by the vertical green line on Figure \ref{fig:IVG}. At lower gate potentials, the devices enter the Schottky barrier regime where the current is conducted as a mixture of thermionic emission and tunneling of the holes. The threshold potential at $V_G=V_{TH}=-2.7$ V is defined as the point where the valence band edge rises above the chemical potential of the source and has been identified by calculating the projected DOS. In this condition, we find ON-currents of 1.6, 0.79 and 1.2 nA/$\mu$m, respectively. Previous experimental measurements on \ce{MoTe2} 2H-1T' heterophase devices with channel thickness between 5 and 8 nm and channel lengths on the $\mu$m scale have presented ON-currents around 15 nA/$\mu$m at a bias of 50 mV and 39.6 nA/$\mu$m at a bias of 0.1 V.\cite{Sung2017,Ma2019} By non self-consistently increasing the bias, we can compare our ON-currents with these values and get a reasonable agreement. The extrapolated ON-currents of the 20 nm device are summarized in Table \ref{tab:I_comp} and the extrapolated ON-currents of the two other devices can be found in the Supplementary Information.$^\dag$ In the table, we also include the measurements on the optimized device fabricated by Zhang \emph{et al.}\cite{Zhang2019apr} which results in an ON-current which is two orders of magnitude higher than the other measurements and our results.

The energy efficiency of the devices is reflected both in the value of the sub-threshold slope and in the power supply voltage, $V_{DD}$. We define $V_{DD}$ as the gate potential difference between the threshold potential and the potential of minimal current. This value is related to the Schottky barrier height since a lower barrier at the source will result in a lower flat-band potential and decrease this value. The 10 nm device has a $V_{DD}$ value of 5.7 V and the 18 and 20 nm devices have $V_{DD}$ values of 5.1 V. These values are around an order of magnitude higher than what typical silicon based devices require.\cite{IRDS2020} The power supply voltage required in the 10 nm device is larger than in the two other devices due to the proximity of the contacts. The interface dipole fields at the source and drain interacts which results in a less efficient gate response and a larger flat-band potential. The effect is illustrated in the Supplementary Information.$^\dag$ In the 10 nm device, also the OFF-current is larger than in the other devices since this device isn't fully turned off when the switch in polarity occurs. It is important to note that the switch in polarity is directly related to the bandgap of the channel. PBE systematically underestimates the bandgap of the 2D TMDs which means that the potential of the polarity switch is underestimated as well.

\begin{figure*}
    \centering
    \includegraphics[width=.99\textwidth]{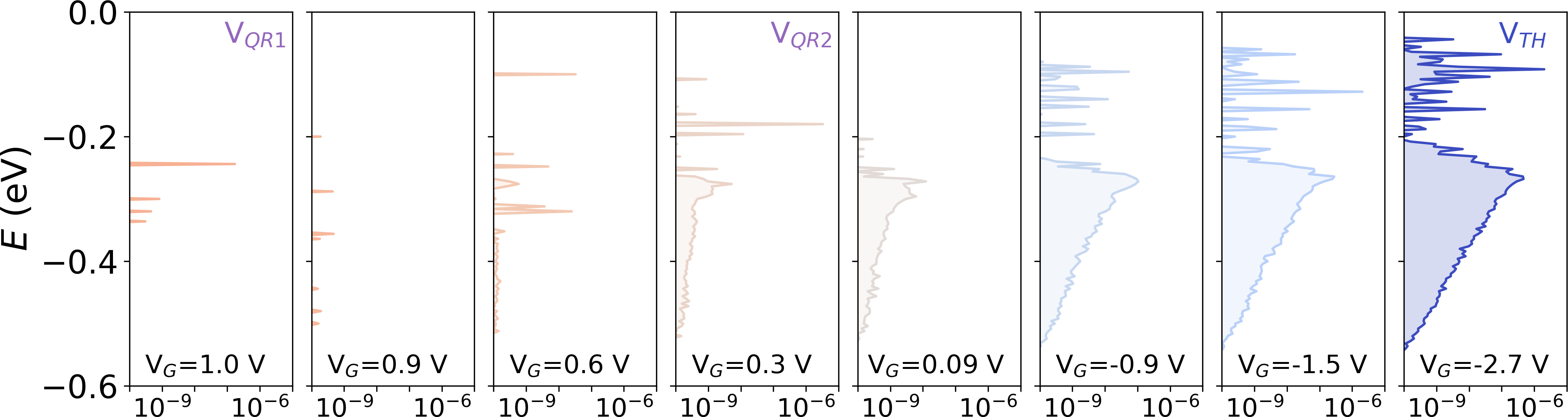}
    \caption{Spectral current in units of q/h in the transistor with a channel length of 20 nm. The two large current peaks marked by the purple stars on Figure \ref{fig:IVG} are denoted QR1 and QR2. At the quantum resonances (QR), significant contributions in the tunneling current occur through these peaks.}
    \label{fig:spectral}
\end{figure*}

The Schottky barrier regime between $V_{TH}$ and $V_{FB}$ is dominated by large spikes in the current which have been marked by the purple stars on Figure \ref{fig:IVG}. The spikes in the current occur due to resonant tunneling and the effect is especially pronounced in the 20 nm device. On Figure \ref{fig:spectral}, the spectral current of this device is shown at different gate potentials between the first purple star at $V_G=V_{QR1}=1.0$ V and the threshold potential. As the device is turned on, the peaks associated with the quantized states in the channel appear and disappear whenever a quantized state contribute to the tunneling current. At the two potentials corresponding to the purples stars, large peaks due to resonant tunneling can be identified at low energies.

\begin{figure}
    \centering
    \includegraphics[width=.99\columnwidth]{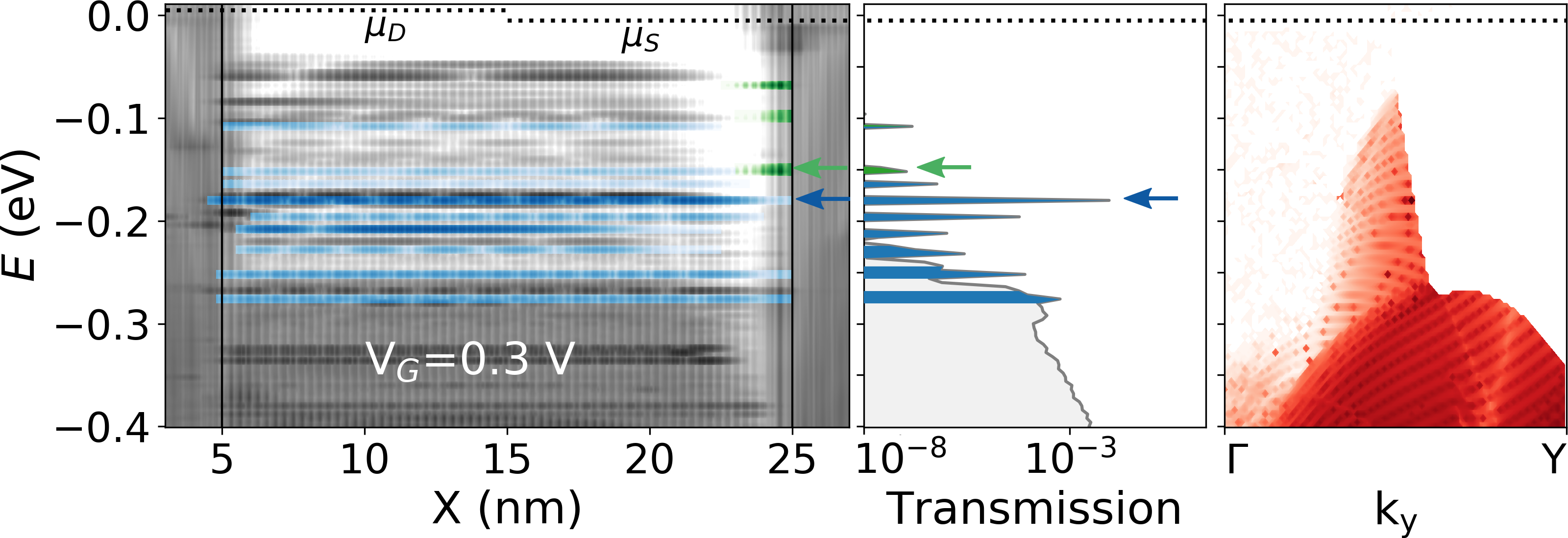}
    \caption{Quantum effects in \ce{MoTe2} heterophase transistors. \textbf{a} shows the projected DOS of the 20 nm transistor at a gate potential of $V_G=0.3$ V. The chemical potential of the drain, $\mu_D$, and the source $\mu_S$ are indicated by the dotted black lines. The interface states are marked by green contours and some of the standing waves in blue contours. \textbf{b} shows the transmission spectrum where the peaks due to the standing waves and interface states have been marked in blue and green. \textbf{c} shows the k-resolved transmission perpendicular to the transport direction. The most predominant peak in the transmission spectrum is marked by the blue arrows.}
    \label{fig:quantum}
\end{figure}

The effect is most pronounced at a gate potential of 0.3 V. On Figure \ref{fig:quantum}a and \ref{fig:quantum}b, the projected DOS at this gate potential is shown along with the transmission spectrum. Two kinds of quantum phenomena cause the peaks to occur in the transmission and spectral current. The first type is the interface states discussed in the previous chapter which have been highlighted in green on the figure, the second type is standing waves in the channel which form due to quantum confinement. These have been marked in blue. In these devices, extraordinary large peaks occur in the transmission when a standing wave bridge the Schottky barrier. An example of this is highlighted by the blue arrows on the figure.

Note, that the transmission is quite low even below the bottom of the barrier. This is due to the restriction of momentum conservation in ballistic transport. In p-type \ce{MoTe2} heterophase contacts, the ballistic transport is significantly reduced since only very few k-points along the $Y \rightarrow \Gamma$ path are occupied both in the metallic and semiconducting phase. Conserving the momentum perpendicular to the transport direction is therefore only possible for a narrow range of k-points. This is illustrated in Figure \ref{fig:quantum}c, where the $k_y$-dependence of the transmission is illustrated. We therefore expect that including the effects of electron phonon scattering might actually increase the device performance. 

To summarize, we find that resonant tunneling and momentum conservation play a very large role for the device performance even in these relatively long channel devices. Comparing with the behavior of the Schottky contacts, we find that the standing waves in the channel affect the current to a larger degree than the effect of the interface resonances. Since the standing waves are a property of the semiconductor and not the metal, the electrostatic field from the gate moves these significantly up and down in energy. When a standing wave bridges the Schottky barrier, the current increases dramatically. This results in large current fluctuations in the transfer characteristics with peaks in the current rising more than an order of magnitude. As it was demonstrates on Figure \ref{fig:quantum}a and \ref{fig:quantum}b, the interface states also contribute to the current in the transistors but the effect is smaller and relatively constant with the gate potential since they have origin in the metallic phase.

\section{Discussion\label{sec:disc}}

Previous calculations on similar devices have been performed by Fan \emph{et al.}\cite{Fan2018} These devices differ from ours since the T phase rather than T' phase was used as contacts, the channel lengths were much smaller, between 5 and 8 nm, the interface was along the armchair direction, and the devices had both a top and back gate with EOT's of only 0.5 nm. Furthermore, that study used the Local Density Approximation (LDA) functional rather than the PBE functional. The results from those calculations showed ON-currents which are around an order of magnitude larger than those reported here and a power supply voltage around 1 V. For \ce{WTe2} and \ce{WSe2} based devices, the results from Fan \emph{et al.} suggest that devices based on the T phase generally conduct more current than T'-based devices. The reason for this difference might lie in the difference in dispersion between the T and T' phase. Since the T phase is metallic whereas the T' phase is semi-metallic, the states are generally more densely distributed in k-space which allows for a greater probability of ballistic transport of carriers. Whether the rest of the difference in the ON-current results from the difference in the device setup or in the method between this study and ours is difficult to say. However, the behavior of the interfaces along the zigzag direction seems most relevant since the experimental evidence suggests that the interfaces occur along this direction.\cite{Sung2017,Xu2020,Han2020}

In Table \ref{tab:I_comp}, we have summarized our results and compared to the three experimental studies mentioned in the previous section and to the International Roadmap for Devices and Systems (IRDS) requirements\cite{IRDS2020} of a 2022 high-performance logic device. In summary, our 18 and 20 nm devices show a performance which is comparable to previous measurements on devices on top of \ce{SiO2} substrates. Comparing to the IRDS goal, the energy efficiency and total ON-current of the devices are not competitive with the performance of silicon based devices. The experimental results by Zhang \emph{et al.}\cite{Zhang2019apr}, on the other hand, suggests that the IRDS requirements might be fulfilled by using a high-$\kappa$ dielectric as the gate oxide which is also the common practice in silicon based devices. However, the device dimensions in the study by Zhang \emph{et al.}~is much larger than both our calculations and the IRDS requirements. This therefore opens the question of whether the heterophase design will be competitive when scaling down the channel length of the devices. Judging from these considerations, the idealised monolayer limit with perfect interfaces might not be preferable since ballistic transport through such geometries are limited by momentum conservation. However, electron-phonon scattering might remove some of this effect and since the device performance is dominated by the local atomic configuration at the interface, the presence of defects in the interface might also dramatically alter the behavior. Furthermore, as demonstrated for the Schottky contacts, a high carrier density will smear out the difference between interface geometries and result in a relatively low Schottky barrier. A heavy doping of the H phase might therefore be a viable approach to obtain a competitive design.

\section{Conclusion\label{sec:conc}}

Our simulations of the heterophase FET devices demonstrated that quantum effects play a large role in the performance of these devices even at a channel length of 20 nm's. We find that quantum resonances result in resonant tunneling and large fluctuations in the current in the Schottky barrier regime of the transfer characteristics. The quantum resonances stem from both interface states at the contacts and standing waves in the channel. The interface states are a material specific property of the devices whereas the standing waves are an effect of the scale and dimensionality of the device and can be expected to be a general property of 2D transistors. Furthermore, we find that the restriction of momentum conservation inhibit the ballistic transport significantly in these idealised devices. 

All-in-all, our results demonstrate many challenges related to the heterophase FET design. Effects from electron-phonon scattering and interface defects might actually improve the performance and calculations estimating such effects could shed light on this possibility. Likewise, moving from the monolayer to the few layer limit or ensuring a high carrier density in the channel might resolve some of the difficulties since this will somewhat smear out some of the dependency on the interface geometry. The results by Nourbakhsh \emph{et al.}\cite{Nourbakhsh2016} where \ce{MoS2} based heterophase devices with both one and three layers were investigated showed that the 3-layer device performs better, which agrees with this conclusion.

\begin{acknowledgments}
This work is partly funded by the Innovation Fund Denmark (IFD) under File No. 5189-00082B. We would like to thank Kurt Stokbro and Daniele Stradi for their contributions.
\end{acknowledgments}

\bibliography{refs}

\end{document}